\documentclass[12pt,preprint]{aastex}
\begin{document}
\title{The Nonlinear Evolution of Galaxy Intrinsic Alignments}
\author{Jounghun Lee}
\affil{Department of Physics and Astronomy, FPRD, Seoul National University, 
Seoul 151-747, Korea} 
\email{jounghun@astro.snu.ac.kr}
\author{Ue-Li Pen}
\affil{Canadian Institute for Theoretical Astrophysics, Toronto, ON M5S 3H8, 
Canada} 
\email{pen@cita.utoronto.ca}

\begin{abstract}
The non-Gaussian contribution to the intrinsic halo spin alignments 
is analytically modeled and numerically detected. Assuming that the growth 
of non-Gaussianity in the density fluctuations caused the tidal field to have 
nonlinear-order effect on the orientations of the halo angular momentum, 
we model the intrinsic halo spin alignments as a linear scaling of the 
density correlations on large scales, which is different from the previous 
quadratic-scaling model based on the linear tidal torque theory.
Then, we analyze the halo catalogs from the recent high-resolution 
Millennium Run simulation at four different redshifts ($z=0,0.5,1$ and $2$) 
and measure quantitatively the degree of the nonlinear effect on the halo 
spin alignments and its changes with redshifts. A clear signal of spin 
correlations is found on scales as large as $10h^{-1}$Mpc at $z=0$, which 
marks a detection of the nonlinear tidal effect on the intrinsic halo 
alignments. We also investigate how the nonlinear effect depends on the 
intrinsic properties of the halos. It is found that the degree of the 
nonlinear tidal effect increases as the halo mass scale decreases, the 
halo specific angular momentum increases, and the halo peculiar velocity 
decreases. We discuss implication of our result on the weak gravitational 
lensing. 
\end{abstract}

\section{INTRODUCTION}

The intrinsic galaxy alignment refers to a cosmic phenomenon that the 
orientations of the galaxy spin axes (or galaxy ellipticities) are not 
random but locally aligned between neighbors due to the spatial correlations 
of the initial tidal field. The topic of the intrinsic galaxy alignments has 
recently drawn considerable attention in the field of the large-scale 
structure for three reasons. Firstly, since it is believed to be induced by 
the spatial correlations of the initial tidal field, it will be possible in 
principle by measuring the galaxy intrinsic alignments to reconstruct the 
initial density field on galactic scales 
\citep{lee-pen00,lee-pen01,lee-pen02,lee-erd07}.

Secondly, the initial tidal correlations are also responsible for the web-like 
distribution of the large scale structures \citep{bon-etal96}. Thus, the galaxy 
intrinsic alignments will be useful to characterize and quantify the influence 
of the large-scale structures on the orientations of the galaxies distributed 
in a cosmic web \citep{nav-etal04,tru-etal06,pat-etal06,ara-etal07,hah-etal07}.
 
Thirdly and most importantly, the galaxy intrinsic alignments could cause 
non-negligible contamination on weak lensing signals. Plenty of efforts have 
been made so far to distinguish and separate the intrinsic contaminations 
from the galaxy extrinsic alignments caused by the weak gravitational lensing 
effect \citep{cro-met00,hea-etal00,cat-etal01,cri-etal01,jin02,hui-zha02,
hey-hea03,tak-whi04,hir-sel04,kin05,man-etal06}.

For these reasons, it is of fundamental importance to predict the degree 
of galaxy intrinsic alignments with a valid theory and measure quantitatively 
their signals in practice. In fact, \citet{pen-etal00} have provided an 
analytic model for the intrinsic galaxy spin alignments in the frame of the 
linear tidal torque theory \citep{dor70,whi84,cat-the96}. According to their  
model, the galaxy intrinsic alignments can be expressed in terms of a 
quadratic scaling of the density correlation function and thus 
are expected to exist only at small distances of order of a few Mpc. 
A crucial implication of their analytic model is that due to the quadratic 
scaling nature of the galaxy intrinsic alignments the cross-correlations 
between the gravitational lensing shears from large scale density field 
and the galaxy intrinsic alignments with local tidal shears will be zero.

\citet{pen-etal00} tested their analytic predictions against low-redshift 
samples of spiral galaxies and found that the observed signals are 
consistent with the analytic predictions. Yet, their detections 
suffered from poor-number statistics and thus were still tentative. Later, 
several other authors have confirmed the existence of galaxy intrinsic 
alignments at small scales either in numerical simulations or in low-redshift 
observational surveys \citep{hea-etal00,jin02,bro-etal02,hey-etal04}.

However, \citet{hui-zha02} has pointed out that it might be invalid to 
describe the galaxy intrinsic alignments on large scales as a quadratic 
function of the density correlation. Their logic is as follows: 
Since the quadratic scaling is based on the linear tidal torque theory 
which adopts a somewhat oversimplified  assumption that the tidal field is 
Gaussian in the subsequent evolutionary stages, it should not be a good 
approximation to describe the galaxy intrinsic alignments  on large scales. 
In reality, the density fluctuations will develop non-Gaussianity via 
gravity which would in turn lead to non-negligible contributions of the 
nonlinear-order of the tidal tensors to the generation of the galaxy 
angular momentum. Due to this nonlinear effect on the galaxy angular 
momentum, their intrinsic spin alignments should be better approximated 
as linear scaling with the density correlation function. Because the 
linear scaling of the density correlation drops much slowly than the 
quadratic scaling, the intrinsic spin alignments would not be completely 
negligible even on large scales. If their claims turn out to be true, then 
it will have a significant impact not only on the weak lensing analysis but 
also on our fundamental understanding of the evolution of the tidal alignments. 

In the light of their claims, the following questions naturally arise; 
Does the nonlinear tidal effect on the galaxy intrinsic alignments really 
exist to a non negligible degree? If so, at what epochs and on which scales 
does its contribution begin to be significant? Does it depend on the 
intrinsic properties of the galaxies or dark halos? Our goal here is to 
answer the above questions using both analytical and numerical methods.

The outline of this paper is as follows: \S 2, we overview briefly the 
previous analytic model for the galaxy spin-spin alignments based on the 
linear tidal torque theory and propose a new model to account for the 
nonlinear tidal effect. \S 3, we report a detection of the signals 
of the nonlinear tidal effect on the intrinsic spin alignments of dark 
matter halos simulated in N-body experiments and show how the signals 
depend on redshift, scale, and the halo intrinsic property. \S 4, we 
summarize our results and discuss the implications of our work on 
the weak lensing effect.

\section{ANALYTIC MODEL}

\subsection{Overview of the Linear Tidal Torque Theory}

The linear tidal torque theory explains that unless a proto-halo has a 
perfectly spherical shape, it can acquire spin angular momentum at first 
order through its tidal interaction with the surrounding matter 
\citep{dor70,whi84,cat-the96}. The main prediction of the linear tidal 
torque theory is that the proto-halo angular momentum vector, ($J_{i}$), 
is proportional to the anti-symmetric product of the two tensors, the 
inertial momentum tensor $(I_{ij})$ and the local tidal shear tensor,
$(T_{ij})$, as
\begin{equation}
\label{eqn:ang}
L_{i} \propto \epsilon_{ijk}T_{jl}I_{lk}. 
\end{equation}
Here the defintions of $I_{ij}$ and $T_{ij}$ are given as
\begin{equation}
\label{eqn:int}
I_{ij} \equiv \int_{V}\! d^{3}{\bf q}\rho({\bf q})q_{i}q_{j}, \qquad
T_{ij}\equiv \frac{\partial^{2}}{\partial q_{i}\partial q_{j}}\Phi({\bf q}).
\end{equation}
where ${\bf q}$ is the Lagrangian position of the particles that reside in 
the proto-halo regions, $V$ and $\rho({\bf q})$ are the Lagrangian volume 
and the density of the proto-halo region, respectively, and $\Phi({\bf q})$ 
is the velocity perturbation potential.

Given the property of the perfectly anti-symmetric tensor 
$\epsilon_{ijk}$ in equation (\ref{eqn:ang}), an additional 
condition has to be satisfied for the first-order generation of the 
proto-halo angular momentum: the principal axes of $I_{ij}$ and $T_{ij}$ 
have to be misaligned with each other \citep{cat-the96}. 
Furthermore, a crucial implication of equation (\ref{eqn:ang}) is 
that if this additional condition is satisfied and thus the angular 
momentum of a proto-halo is generated at first order, then the direction 
of the proto-halo angular momentum is not random but preferentially 
aligned with the principal axes of the local tidal tensor \citep{lee-pen00}. 

Numerical experiments have revealed that the principal axes of $(I_{ij})$ 
and $(T_{ij})$ are correlated strongly but not perfectly 
\citep{lee-pen00,por-etal02}, which indicates that the first-order generation
of the proto-halo angular momentum is not so efficient and the degree of the 
alignments between the proto-halo spin directions and the principal axes of 
the local tidal shear tensors would not be so high.

Motivated by this numerical clue, \citet{lee-pen00} suggested the following 
quadratic formula to quantify the expected degree of the alignments between 
the halo spins and the local tidal shears, generalizing the linear tidal 
torque theory:
\begin{equation}
\label{eqn:stc}
\langle\hat{L}_{i}\hat{L}_{j}\rangle = \frac{1+a}{3}\delta_{ij} -
a\hat{T}_{ik}\hat{T}_{kj}.
\end{equation}
where $(\hat{L}_{i})$ is the unit spin vector of a halo, $(\hat{T}_{ij})$ is 
the unit traceless tidal tensor smoothed on the halo mass scale, and $a$ is 
a correlation parameter in the range of $[0,3/5]$ which measures the strength 
of the correlation between $(\hat{L}_{i})$ and $(\hat{T}_{ij})$. It is 
worth recalling the fact that the ensemble average  
$\langle\hat{L}_{i}\hat{L}_{j}\rangle$ in the left hand side of equation 
(\ref{eqn:stc}) is obtained by taking the average of $\hat{L}_{i}\hat{L}_{j}$ 
from the sample halos having all different mass, while the tidal shear field 
in the right hand side of equation (\ref{eqn:stc}) is smoothed on the single 
mass scale which amounts to the mean halo mass (i.e., the mean value of the 
mass of the sample halos).\citet{lee-pen00} derived equation (\ref{eqn:stc}) 
empirically, relating the angular momenta of the halos with different mass 
to the tidal shear field smoothed on the mean halo mass.

The correlation parameter $a$ in equation (\ref{eqn:stc}) which characterizes 
equation (\ref{eqn:stc}) is introduced to take into account strong 
correlations between $(I_{ij})$ and $(T_{ij})$ and any modification in the 
subsequent evolution after the moment of the turn-around. 
The two extreme cases of $a=0$ and $a=3/5$ represent no correlation and 
maximum correlation between $(\hat{L}_{i})$ and $(\hat{T}_{ij})$, 
respectively. It is expected that the true value of $a$ would be between 
these two extremes.

The linear tidal torque theory itself cannot make any prediction on the 
value of $a$. Thus, the true value of this free parameter $a$ has to be 
determined empirically. So far, several attempts have been made to measure 
the value of $a$ either from numerical simulations or from observations 
\citep{lee-pen00,por-etal02,nav-etal04,tru-etal06,lee-erd07}. Although no 
general consensus has been reached on the true value of $a$, most of the 
numerical and observational evidences indicated that the true value of 
$a$ deviates from zero.

\subsection{The First Order Approximation to the Halo Spin Correlations}

A difficulty in constraining the value of $a$ lies in the fact that 
it is hard to measure the principal axes of the linear tidal field in 
real space.  An alternative approach to the true value of $a$ is to 
measure the intrinsic spin-spin correlations of dark halos \citep{pen-etal00}.
The local alignments between the halo spin vectors and the tidal shear tensors 
will lead to spatial spin alignments between the neighbor halos. 
Using equation (\ref{eqn:stc}), \citet{pen-etal00} derived the following 
analytic model for the halo spin-spin correlations:
\begin{equation}
\label{eqn:1st}
\eta(r) \equiv \langle\vert\hat{\bf J}({\bf x})\cdot
\hat{\bf J}({\bf x + r})\vert^{2}\rangle - \frac{1}{3}
\approx \frac{a^2}{6}\frac{\xi^{2}(r;R)}{\xi^{2}(0;R)}.
\end{equation}
Here, the constant $1/3$ represents the value of 
$\langle\vert\hat{\bf J}({\bf x})\cdot\hat{\bf J}({\bf x}+{\bf r})
\vert^{2}\rangle$ for the case of no alignment, and  $\xi(r;R)$ is the linear 
two-point correlation function defined as
\begin{equation}
\label{eqn:xi}
\xi(r;R) \equiv \int_{-\infty}^{\infty}\Delta^{2}(k)\frac{\sin kr}{kr}
W^{2}(kR)d\ln k,
\end{equation}
where $\Delta^{2}(k)$ is the dimensionless power spectrum and $W(kR)$ 
is the top-hat spherical filter of scale radius, $R$, which is related 
to the halo mass scale $M$ as $R \equiv [3M/(4\pi\bar{\rho})]^{1/3}$ with 
the mean density $\bar{\rho}$. 

Note that equation (\ref{eqn:1st}) is proportional to the square of the 
two point density correlation $\xi(r)$, which means that the halo spin-spin 
correlations would decrease very rapidly. For instance, on a galactic scale 
($R \sim 1 h^{-1}$Mpc), it is expected that the halo spin-spin 
correlations exist only at distances of order of a few Mpcs and 
effectively vanishes at larger distances.

Since the tidal field on different scales are cross-correlated, equation 
(\ref{eqn:stc}) leads to the existence of the cross-correlations of the 
spin axes of halos on different mass scales. In accordance with equation 
(\ref{eqn:1st}) the halo spin cross-correlations on different 
scales ($R_{1}$ and $R_{2}$) are modeled as
\begin{equation}
\label{eqn:crr}
\eta_{C}(r) \approx \frac{a^2}{6}\frac{\xi^{2}_{C}(r;R_{1},R_{2})}
{\xi^{2}_{C}(0;R_{1},R_{2})}.
\end{equation}
where $\xi^{2}_{C}(r;R_{1},R_{2})$ is the density cross-correlation 
function defined as 
\begin{equation}
\label{eqn:cxi}
\xi_{C}(r;R_{1},R_{2}) \equiv 
\int_{-\infty}^{\infty}\Delta^{2}(k)\frac{\sin kr}{kr}
W(kR_{1})W(kR_{2})d\ln k.
\end{equation} 
It is worth mentioning that the value of $a$ for the cross correlation 
$\eta_{C}(r)$ would not be same as its value for the auto-correlation 
$\eta(r)$ in equation (\ref{eqn:1st}) due to the difference in the smoothing 
scale for the tidal field. Note that equation (\ref{eqn:crr}) is a modified 
version of equation (\ref{eqn:1st}), both of which are based on equation 
(\ref{eqn:stc}). The correlation parameter $a$ in equation (\ref{eqn:1st}) 
represents the spin-spin correlations caused by the spatial correlations of 
the tidal fields smoothed on the same scale (one mean halo mass scale, $R$). 
While the correlation parameter $a$ in equation (6) represents the spin-spin 
correlations caused by the spatial correlations of the tidal fields on two 
different scales (two mean halo mass scales, $R_{1}$ and $R_{2}$). 
Since the spatial correlations of the two tidal fields smoothed on two 
different scales, $R_{1}$ and $R_{2}$ (say, $R_{1} < R_{2}$), are lower 
than the case that the two tidal fields are smoothed on the same scale 
of $R_{2}$ and higher than the case that they are smoothed on the same scale 
of $R_{1}$, it is naturally expected that the value of $a$ for the 
cross-correlation $\eta_{C}(r)$ should be different from that for the 
correlation $\eta(r)$.

\subsection{The Nonlinear Effect and Large Scale Correlations of Halo Spins}

As mentioned in \S 1, it was \citet{hui-zha02} who first pointed out that 
equation (\ref{eqn:1st}) is valid only if the gravitational tidal 
field is Gaussian. They argued that the non-Gaussianity in the tidal field
should not be negligible in the nonlinear regime and suggested that the 
halo spin-spin alignments should be better described as a linear scaling 
with the density correlation function on sufficiently large scales. 

Inspired by the work of \citet{hui-zha02}, here we suggest a new formula 
to model the halo spin-spin correlations:
\begin{equation}
\label{eqn:2nd}
\eta(r) \approx \frac{a^2_{\rm l}}{6}\frac{\xi^{2}(r;R)}{\xi^{2}(0;R)} + 
\varepsilon_{\rm nl}\frac{\xi(r;R)}{\xi(0;R)},
\end{equation}
where a nonlinear correlation parameter $\varepsilon_{\rm nl}$ is introduced 
to measure the strength of the nonlinear tidal effect on the halo 
spin-spin correlations. In this model the linear correlation parameter is 
notated as $a_{\rm l}$ since its value could be different between the 
two cases that the nonlinear effect is ignored as in equation 
(\ref{eqn:1st}) and taken into account as in (\ref{eqn:2nd}).
 
Equation (\ref{eqn:2nd}) implies that if the nonlinear contribution 
exists, then the halo spin-spin correlations must exist on larger scale 
than predicted by the linear model (eq.[\ref{eqn:1st}]) since it scales 
linearly with the density correlation function in the nonlinear model. 
That is, the nonlinear tidal effect would generate large-scale correlations 
of the halo spin axes. Of course, the values of the nonlinear correlation 
parameter $\varepsilon_{\rm nl}$ as well as the linear correlation parameter 
have to be determined empirically.

The cross-correlation, $\eta_{C}(r)$ can be also modeled in accordance 
with (\ref{eqn:2nd}) as
\begin{equation}
\label{eqn:crn}
\eta_{C}(r) \approx 
\frac{a^2_{\rm l}}{6}\frac{\xi^{2}_{C}(r;R_{1},R_{2})}
{\xi^{2}_{C}(0;R_{1},R_{2})} + 
\varepsilon_{\rm nl}\frac{\xi_{C}(r;R_{1},R_{2})}{\xi_{C}(0;R_{1},R_{2})}.
\end{equation}
It is worth mentioning that the value of $\epsilon_{\rm nl}$ for the cross 
correlation $\eta_{C}(r)$ would not be necessarily lower than its value 
for the auto-correlation $\eta(r)$ in equation (\ref{eqn:2nd}), unlike the 
case of the linear tidal torque model in \S 2.2. In the nonlinear regime 
where the non-Gaussianity in the density field grows, the occurrence of the 
halo merging is quite frequent. This occurrence of the halo merging plays a 
role of transferring the orbital angular momentum generated by the external 
tidal field into the spin angular momentum of a merged halo. In other words, 
the orbital angular momentum of a system composed of small individual halos 
become the spin angular momentum of a large halo formed through merging of 
the small halos. Given that the orbital angular momentum of the system 
before merging is generated by the external tidal field on larger scale 
\citep{vit-etal02}, the transfer of the orbital angular momentum into the 
spin angular momentum creates cross-correlations of the tidal fields between 
different scales, which in turn generates large-scale cross-correlations 
$\eta_{C}$ in the nonlinear regime.

\section{SIGNALS FROM NUMERICAL SIMULATIONS}

\subsection{Numerical Data}

We use the halo catalogs from the Millennium Run Simulation
\footnote{It is now available at http://www.mpa-garching.mpg.de/millennium} 
of $10^{10}$ dark matter particles for a $\Lambda$CDM universe with the 
cosmological parameters given as $\Omega_{m}=0.25$, $h=0.73$, 
$\sigma_{8}=0.9$ and $n_{s}=1$ \citep{spr-etal05}. 
The simulation was performed in a periodic box of linear size of  
$500 h^{-1}$Mpc to follow the evolution of all particles from $z=200$ 
to $z=0$, each of which has a mass of $8.6\times 10^{8}h^{-1}M_{\odot}$.

We analyze the halo catalogs at four different redshifts $z=0$, $0.5$, $1$ and 
$2$ which contain information on the halo mass, comoving position, comoving 
peculiar velocity and the spin vector. At each redshift, we restrict our 
attention only to those halos which have more than $200$ particles for 
better accuracy in the measurement of the halo spin vector. It amounts to 
selecting only halos which are more massive than 
$17.2\times 10^{9}h^{-1}M_{\odot}$.

Basically, we measure the spatial correlations of the spin axes between 
neighbor halos as a function of separation distance and investigate how 
the correlation strength changes with redshift (z), halo mass ($M$), 
velocity ($v$) and specific angular momentum ($l$, angular momentum per mass). 
Then, we compare the numerical results with the linear and the nonlinear 
analytic models described in \S 2 to determine the best-fit values of the 
correlation parameters, $a$, $a_{l}$, and $\varepsilon_{\rm nl}$.

\subsection{Dependence on the Redshift}

\begin{figure}
  \begin{center}
   \plotone{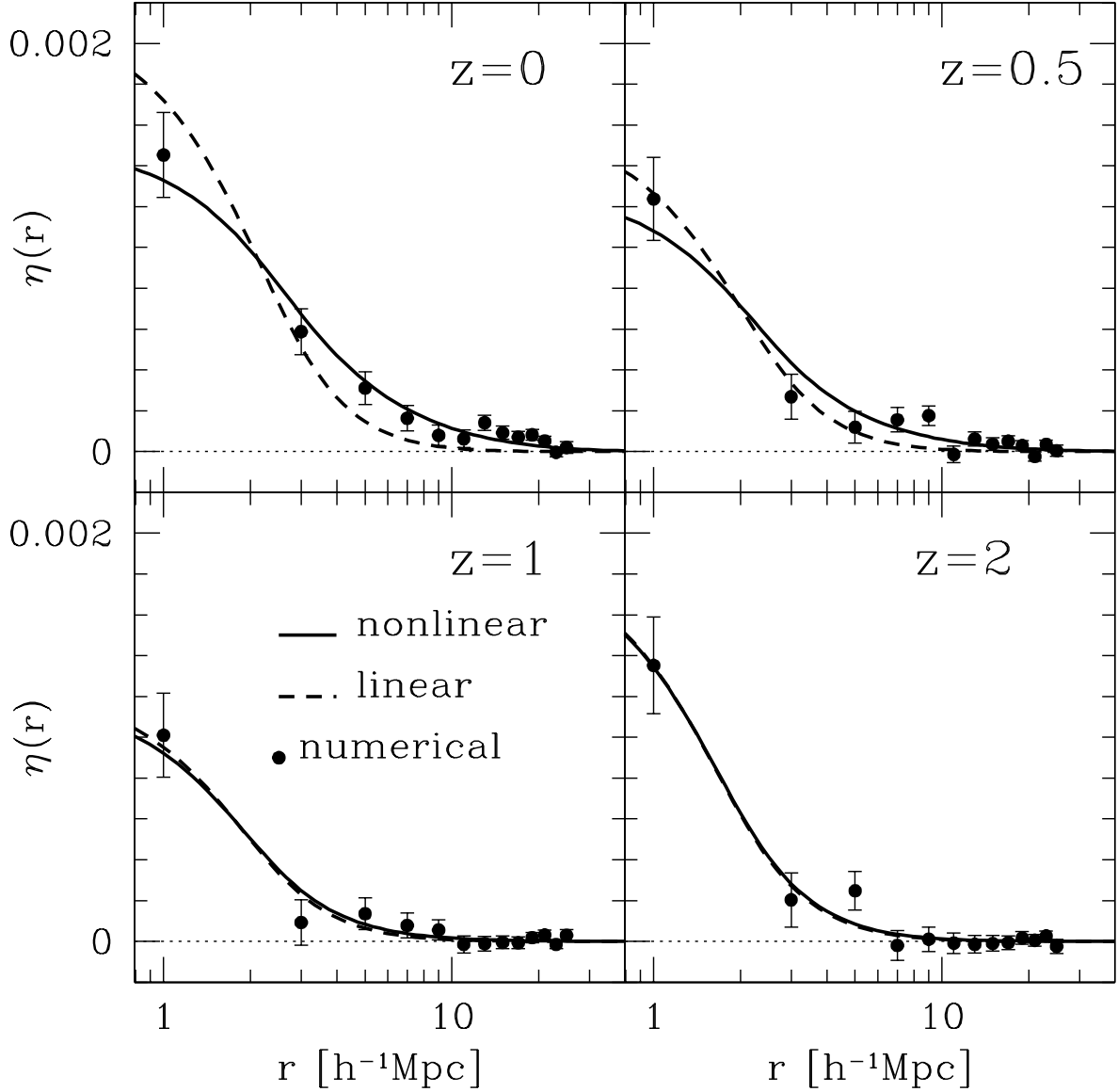}
\caption{Intrinsic halo spin-spin correlations at four different redshifts: 
$z=0,0.5,1$ and $2$ (top-left, top-right, bottom-left, and bottom-right, 
respectively). In each panel, the solid dots represent the numerical 
results, while the solid and dashed lines correspond to the nonlinear 
(eq.[\ref{eqn:1st}]) and linear (eq.[\ref{eqn:1st}]) analytic model, 
respectively.The errors are calculated as the standard deviation for
the case of no correlation. The dotted line represents the case of
no correlation.}
\label{fig:z}
 \end{center}
\end{figure}

Figure \ref{fig:z} plots the halo spin-spin correlation function $\eta(r)$ at 
$z=0$,$0.5$,$1$ and $2$ in the top-left, top-right, bottom-left, and 
bottom-right panel, respectively. In each panel, the solid dots correspond 
to the numerical results from the Millennium data while the solid and the 
dashed lines represent the nonlinear model (eq.[\ref{eqn:2nd}]) and the 
linear model (eq.[\ref{eqn:1st}]), respectively. The case of no alignment 
is also shown as a dotted line in each panel for comparison. 

The errors of the numerical results are calculated as one standard deviation
for the case of no alignment. Basically, it is computed by the formula of
$2/\sqrt{45n_{h}}$ where $n_{h}$ is the number of halos belonging to each bin 
\citep{lee-pen01}. Note that the sizes of the vertical errors in Figure 
\ref{fig:z}  depend on the value of $n_{h}$ which in turn depends on the 
bin size of the spatial separation $r$. Here we choose the bin size of $r$ 
as $2h^{-1}$Mpc for all plots. We have tested whether or not the final 
results depend on the bin size of $r$ and confirmed the robustness 
of the final results.

For the analytic models, the correlation parameters are 
determined through fitting to the numerical results by means of the 
$\chi^{2}$-minimization. For the evaluation of the analytic models, we use 
the formula for the linear power spectrum given by \citet{bar-etal86} 
using the same values of the cosmological parameters that are used for the 
Millennium Run simulations and  the shape parameter $\Gamma = \Omega_{m}h$ 
(private communication with V. Springel).

Table \ref{tab:z} lists the number of halos ($N_{h}$), the mean mass 
($\bar{M}$), the best-fit values of $a$, $a_{nl}$ and $\varepsilon_{\rm nl}$ 
at $z=0-2$. Here, the mean mass $\bar{M}$ for each bin is used to find the 
Lagrangian smoothing scale, $R$ which has to be implemented into 
the analytic models (eqs.\ref{eqn:1st} -\ref{eqn:2nd}).
As shown in Figure \ref{fig:z}, we detect clear signals of the halo spin-spin 
alignments within a few $h^{-1}$Mpc distances at all four redshifts. Note, 
however, a conspicuous difference between the results at low redshifts 
($z=0$ and $0.5$) and at higher redshifts ($z=1$ and $2$). At $z=1$ and $2$, 
the correlation signals rapidly decrease with distance and disappear  
at distances around $10h^{-1}$Mpc. Thus, for the low redshifts, the linear 
model (dashed line) for itself fits the  numerical data pretty well. 
Whereas, at $z=0$ and $0.5$ the correlation signals decrease rather slowly, 
still existent to a nennegligible level at large separations beyond 
$10h^{-1}$Mpc.  Thus, at low redshifts the nonlinear model (solid line) 
fits the numerical data much better than the linear model, predicting the 
existence of the large-scale correlations.

\begin{figure}
  \begin{center}
   \plotone{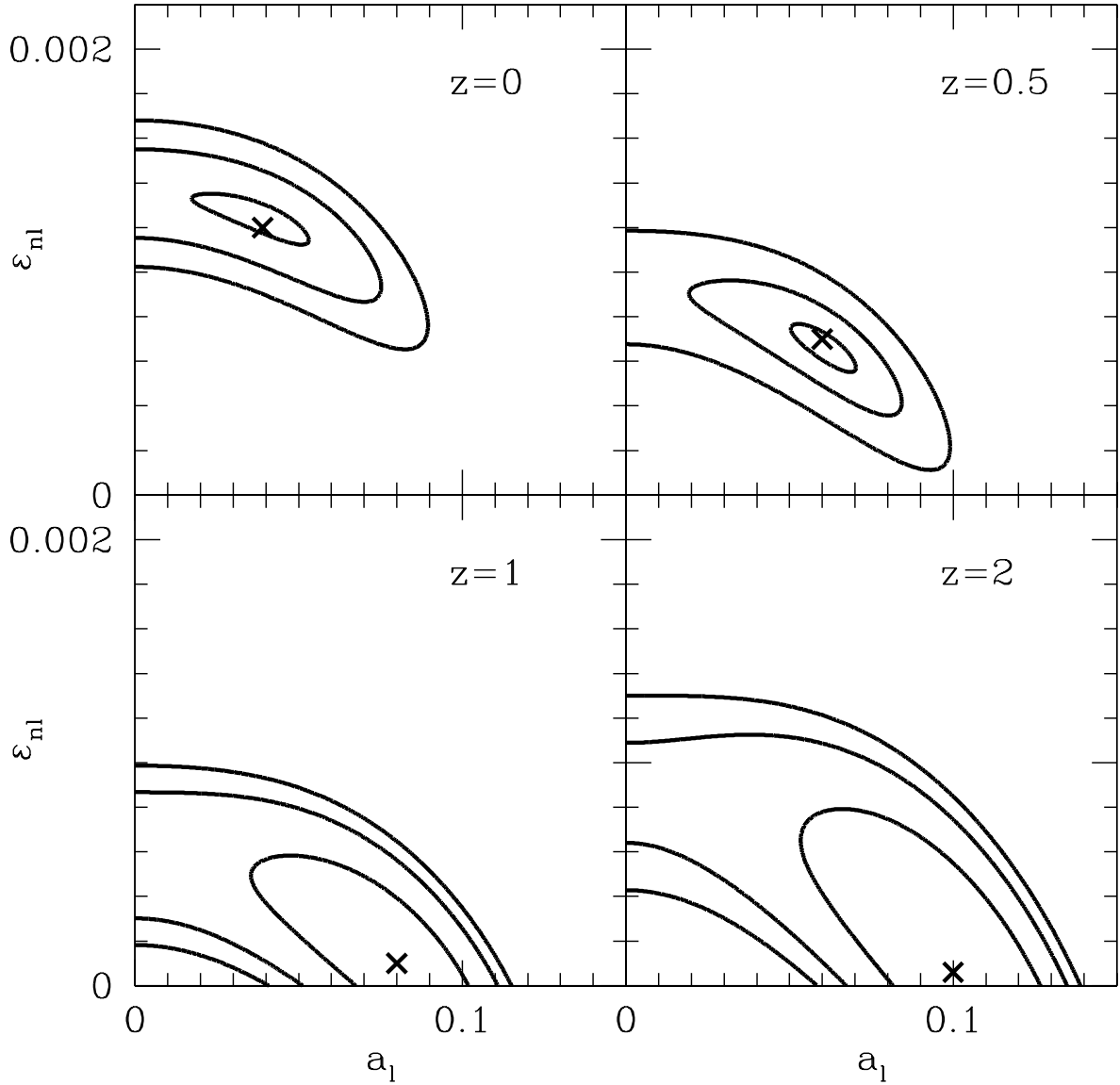}
\caption{$68\%$, $95\%$ and $99\%$ contour plots for the linear and the 
nonlinear correlation parameters, $a_{\rm l}$ and $\varepsilon_{\rm nl}$ 
at $z=0,0.5,1$ and $2$ (top-left,top-right,bottom-left, and bottom-right,
respectively). In each panel, the cross mark represents the best-fit value.}
\label{fig:cz}
 \end{center}
\end{figure}

This phenomena can be quantified in terms of the bestfit-values of 
the correlation parameters listed in Table \ref{tab:z}. At $2$,  
the best-fit value of $\varepsilon_{\rm nl}$ is effectively zero within 
one standard deviation ($\sigma$) while the best-fit value of $a_{\rm l}$ is 
higher than $10\sigma$ (not marginalized level).  In contrast, at $z=0$ the 
value of $\varepsilon_{\rm nl}$ deviates from zero as significantly as 
$12\sigma$ while the value of $a_{\rm l}$ decreases significantly to a 
$3\sigma$ level. It is, however, worthing mentioning here that the 
standard deviation, $\sigma$, here is not marginalized.
Figure \ref{fig:cz} plots the $68\%$, $95\%$ and $99\%$ contours for 
$a_{\rm l}$ and $\varepsilon_{\rm nl}$ at $z=0-2$, showing the degeneracy 
in putting the constraints on the two parameters. As can be seen, at $z=0$ and 
$0.5$, the values of $\varepsilon_{\rm nl}$ deviate from zero at higher than 
$99\%$ level.

Our numerical detection implies that the linear model for the halo spin-spin 
correlations (eq.[\ref{eqn:1st}]) works well at $z > 1$ but the nonlinear 
effect indeed dominates at low redshifts ($z < 1$), producing large-scale 
correlations of the halo spins, which can be well described by our nonlinear 
model (eq.[\ref{eqn:2nd}).

\subsection{Dependence on the Specific Angular Momentum Magnitude}

\begin{figure}
  \begin{center}
   \plotone{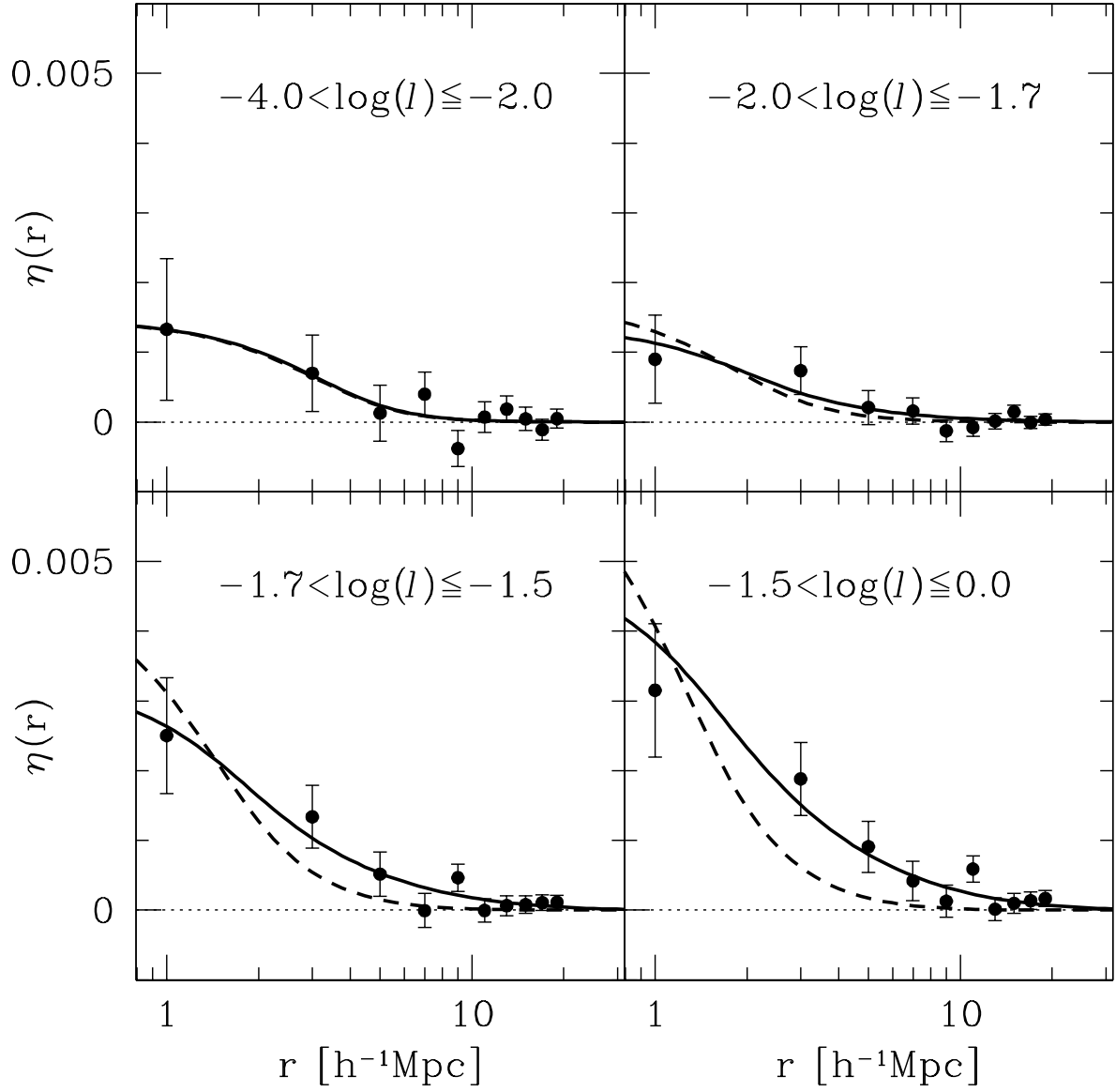}
\caption{Intrinsic halo spin-spin correlations at $z=0$ for the four 
different logarithmic bins of the specific angular momentum magnitude 
$l$ at $z=0$.}
\label{fig:a}
 \end{center}
\end{figure}

\begin{figure}
  \begin{center}
   \plotone{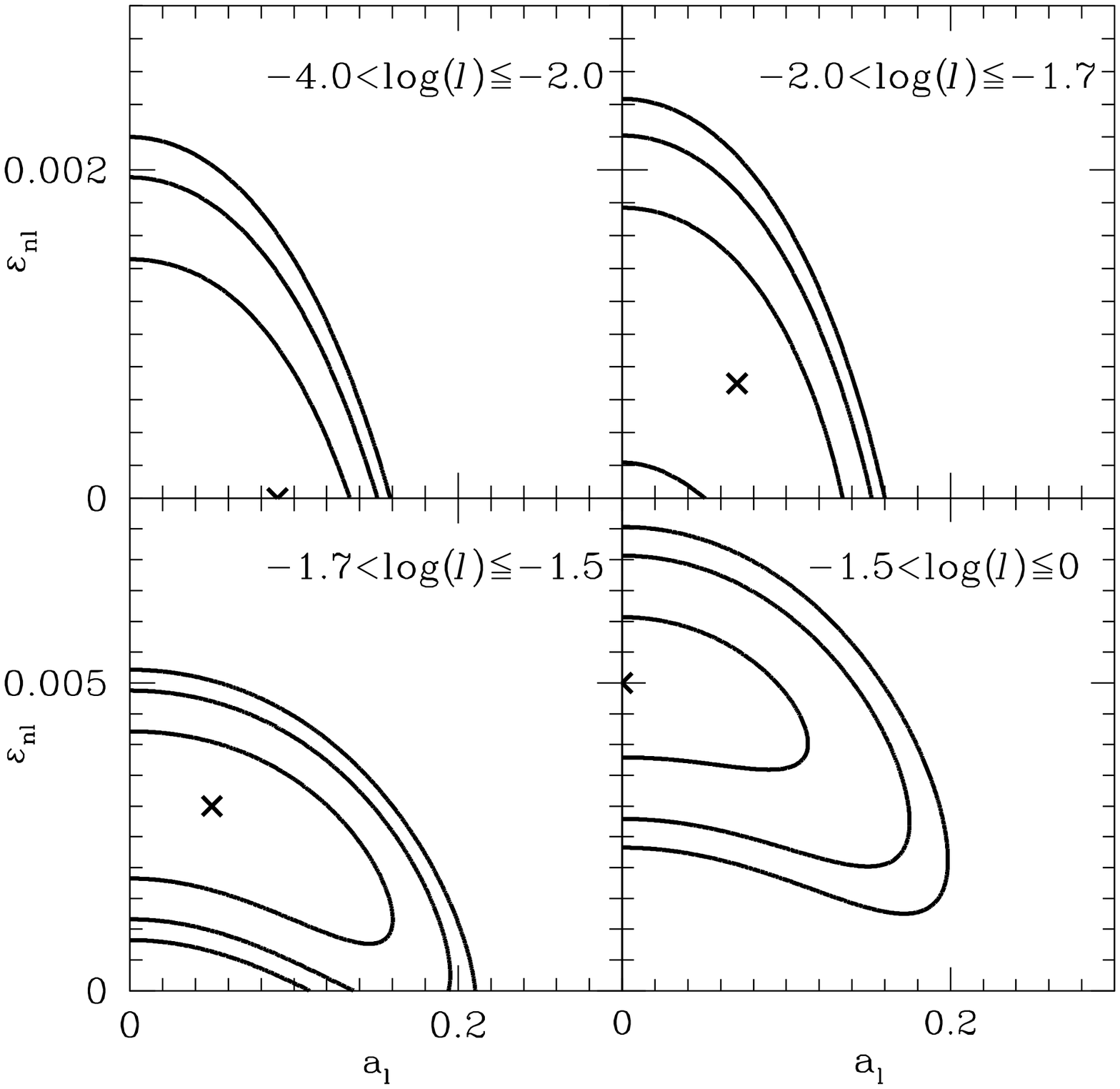}
\caption{$68\%$, $95\%$ and $99\%$ contour plots for $a_{\rm l}$ and 
$\varepsilon_{\rm nl}$at the four logarithmic bins of $l$.}
\label{fig:ca}
 \end{center}
\end{figure}

We define the magnitude of the rescaled specific angular momentum (angular 
momentum per unit mass) as $l \equiv L/M_{*}$ where $L$ and $M_{*}$ 
are the halo angular momentum magnitude and the halo mass in unit of 
$10^{10}h^{-1}M_{\odot}$, respectively. 
To determine quantitatively how the halo spin-spin correlation changes 
with $l$, we classify the halos at  $z=0$ into four logarithmic bins 
of $l$ and measure the correlations separately for each bin. 
Figure \ref{fig:a} plots $\eta(r)$ at four logarithmic bins of $l$ in 
the top-left, top-right, bottom-left, and bottom-right panels, respectively. 
Table \ref{tab:a} also lists the number of halos, the mean mass, the best-fit 
values of $a$, $a_{\rm l}$ and $\varepsilon_{\rm nl}$ for each bin, and 
Figure \ref{fig:ca} plots the $68\%$, $95\%$ and $99\%$ contours for 
$a_{\rm l}$ and$\varepsilon_{\rm nl}$ at the four bins. 

As shown in Figure \ref{fig:a}, in the bins of low-$l$ halos (top two panels), 
the correlations are rather weak and rapidly diminish to zero at separations 
greater than a few $h^{-1}$Mpc. Thus, for this case the linear model with 
the single parameter $a$ fits the data pretty well. In other words, 
for the spin-spin correlations of low-$l$ halos, the values of 
$\varepsilon_{\rm nl}$ are effectively zero as shown in Table \ref{tab:a}. 
While in the bins of high-$l$ halos, the correlations are stronger, still 
existent even at separations as large as $10h^{-1}$Mpc. Therefore for the 
spin-spin correlations of high-$l$ halos, the nonlinear model with non-zero 
value of $\varepsilon_{\rm nl}$ fits the data much better than the linear 
model. Note that for the case of highest-$l$ halos, the value of 
$\varepsilon_{\rm nl}$ is approximately $0.005$, which is five times 
greater than the average value of $0.001$ at $z=0$.

This phenomena that the halos with higher specific angular momentum tend to 
have stronger non-linear (and linear) spin-spin correlations may be explained 
as follows. If a halo is located in a region where the effect of the tidal 
field is very strong, then the halo will possess high specific angular 
momentum. Furthermore, its spin direction will be strongly correlated with 
that of its neighbors since the strong effect of the tidal field tends to 
diminish any randomization of the spin orientations that can occur in the 
subsequent stages.Therefore, those halos which have higher specific angular 
momentum are likely to show strong spin-spin correlations under the strong 
effect of the tidal field. We have also explored whether this phenomena 
occur at higher redshifts $z=0.5$,$1$ and $2$ and found that this is still 
the case.

\subsection{Dependence on the Mass Scale}

\begin{figure}
  \begin{center}
   \plotone{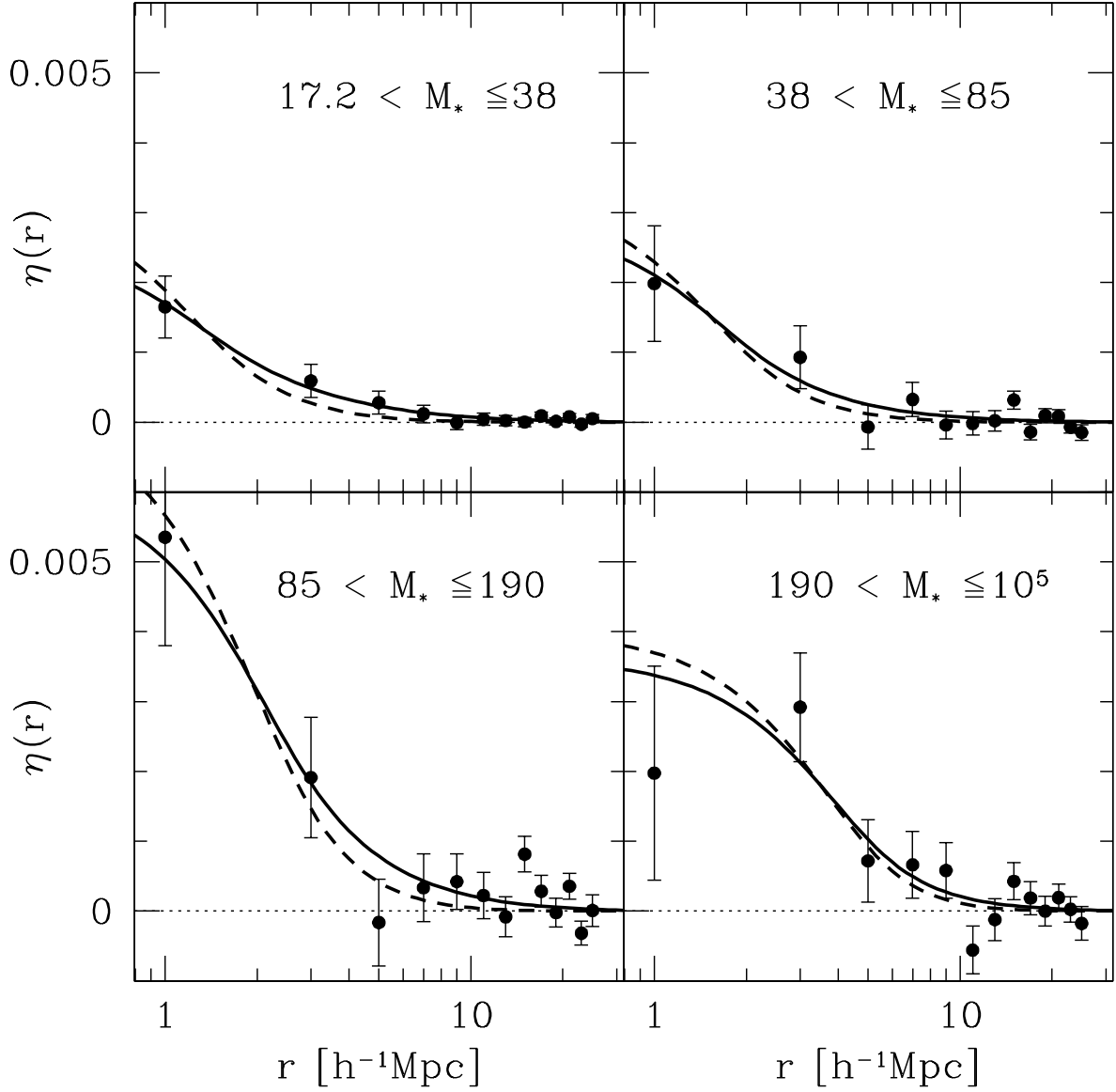}
\caption{The dark halo spin-spin correlations at $z=0$ for the four 
different bins of the halo mass $M_{*}$ in unit of $10^{10}h^{-1}M_{\odot}$.}
\label{fig:m}
 \end{center}
\end{figure}

The halos at  $z=0$ are classified into four bins according to the value 
of the rescaled mass $M_{*}$ and their spin-spin alignments are measured 
separately for each bin. Figure \ref{fig:m} plots $\eta(r)$, 
Table \ref{tab:m} lists the number of halos, the mean mass, the best-fit 
values of $a$, $a_{\rm l}$ and $\varepsilon_{\rm nl}$, and Figure \ref{fig:cz} 
plots the $68\%$, $95\%$ and $99\%$ contours for $a_{\rm l}$ and 
$\varepsilon_{\rm nl}$ for the four bins.

As shown in Figure \ref{fig:m}, in the third bin which includes the galactic 
halos with mass of order of $(1-2)\times 10^{12}h^{-1}M_{\odot}$ 
(bottom-left panel), the correlation signal is strongest,  existent at 
separations as large as $10h^{-1}$Mpc. For this bin, the nonlinear model 
with nonzero value of $\varepsilon_{\rm nl}$ gives a better fit the numerical 
results than the linear with model single parameter $a$. The best-fit value 
of $\varepsilon_{\rm nl}$ deviates from zero by a factor of $3\sigma$ 
as shown in Table \ref{tab:m}. Whereas in the first and the fourth bins 
which include dwarf halos and galaxy-group halos, respectively, the signals 
are rather weak and the correlations disappear at large distances. For 
these bins, the linear model for itself fits the numerical data quite well 
and the values of $\varepsilon_{\rm nl}$ are effectively zero.
This result implies that the strength of the nonlinear effect depends on mass
scales, being strongest on the galactic mass scale. Note, however, here that 
the value of the characteristic mass scale changes with $z$. Thus, the 
results given in \S 3.2 and \S 3.4 are related to each other.

\subsection{Scale Bias and Cross-Correlations}

As mentioned in \S 2.3, if the nonlinear effect dominates on the halo 
spin-spin correlations, then it is likely to produce stronger 
cross-correlations between different mass scales (eq.[\ref{eqn:crn}]). 
We measure the halo spin cross-correlations between the halos of 
different mass bins that are constructed in \S 3.2 (the first bin 
corresponds to the lowest mass scales. see Table \ref{tab:m}), 
and find the best-fit values of $a$, $a_{nl}$ and $\varepsilon_{\rm nl}$ 
by fitting the numerical data to the analytic model (eq.[\ref{eqn:crn}]). 

\begin{figure}
  \begin{center}
   \plotone{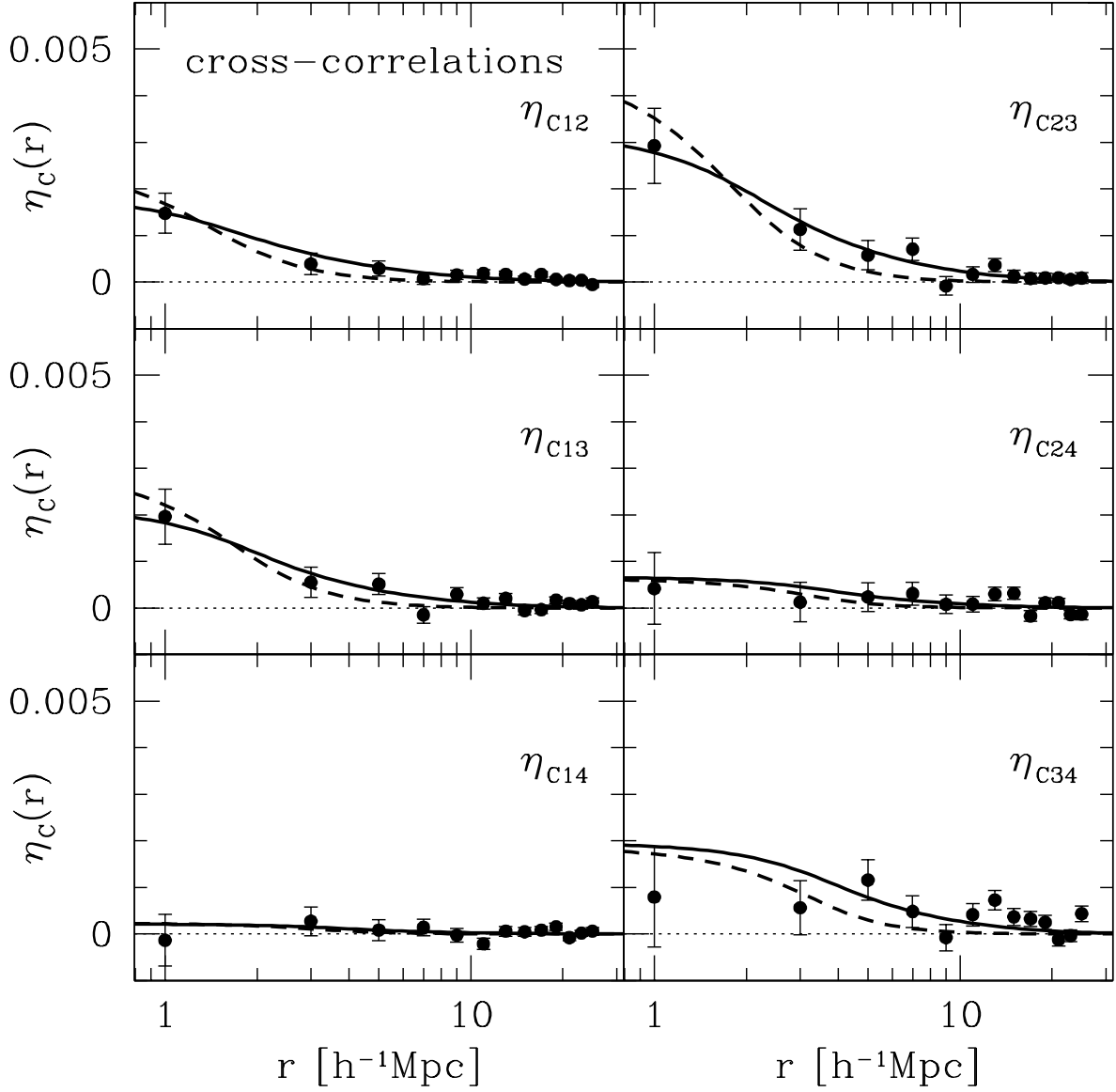}
\caption{Intrinsic halo spin cross correlations between different mass scales 
at $z=0$. In each panel, $\eta_{Cij}$ represents the cross-correlation 
between the $i$-th and the $j$-th mass bin, given in Table \ref{tab:m}. }
\label{fig:b}
 \end{center}
\end{figure}

\begin{figure}
  \begin{center}
   \plotone{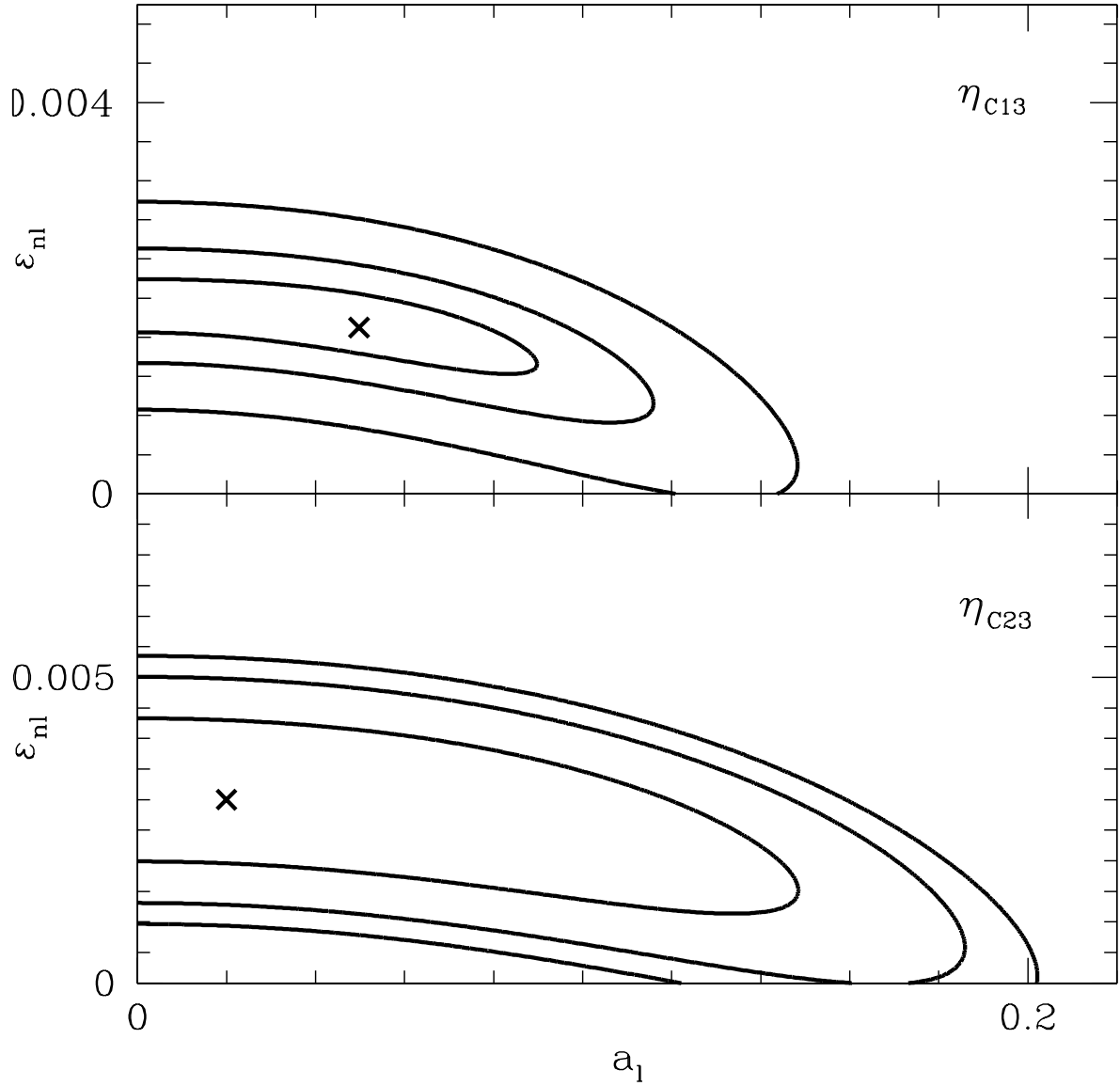}
\caption{$68\%$, $95\%$ and $99\%$ contour plots for $a_{\rm l}$ and 
$\varepsilon_{\rm nl}$ for the spin cross-correlations between halos 
in the $1$st mass bin and the $2nd$ mass bin (top) and between halos 
in the $2$nd mass bin and the $3rd$ mass bin (bottom). }
\label{fig:cb}
 \end{center}
\end{figure}

Figure \ref{fig:b} plots the numerical results of $\eta_{C}$ for the six 
different cases: $\eta_{C12}$, $\eta_{C13}$, $\eta_{C14}$, $\eta_{C23}$, 
$\eta_{C24}$,$\eta_{C34}$, where $\eta_{Cij}$ represents the 
cross-correlations between the $i$-th and the $j$-th mass bins. 
As can be seen, the nonlinear model basically gives better fit to the 
numerical result for all cases, indicating that the nonlinear effect 
increases indeed increase spin cross-correlations between different 
mass scales, as presumed in \S 2.3. Figure \ref{fig:cb} plots the $68\%$, 
$95\%$ and $99\%$ contours for $a_{\rm l}$ and $\varepsilon_{\rm nl}$ 
for these two cases of $\eta_{13}$ and $\eta_{23}$ where the strong signals 
of the cross-correlations are detected. Obviously, the values of 
$\varepsilon_{\rm nl}$ for each case deviates from zero at $95\%$ 
confidence level.

\subsection{Dependence on the Velocity Magnitude}

\begin{figure}
  \begin{center}
   \plotone{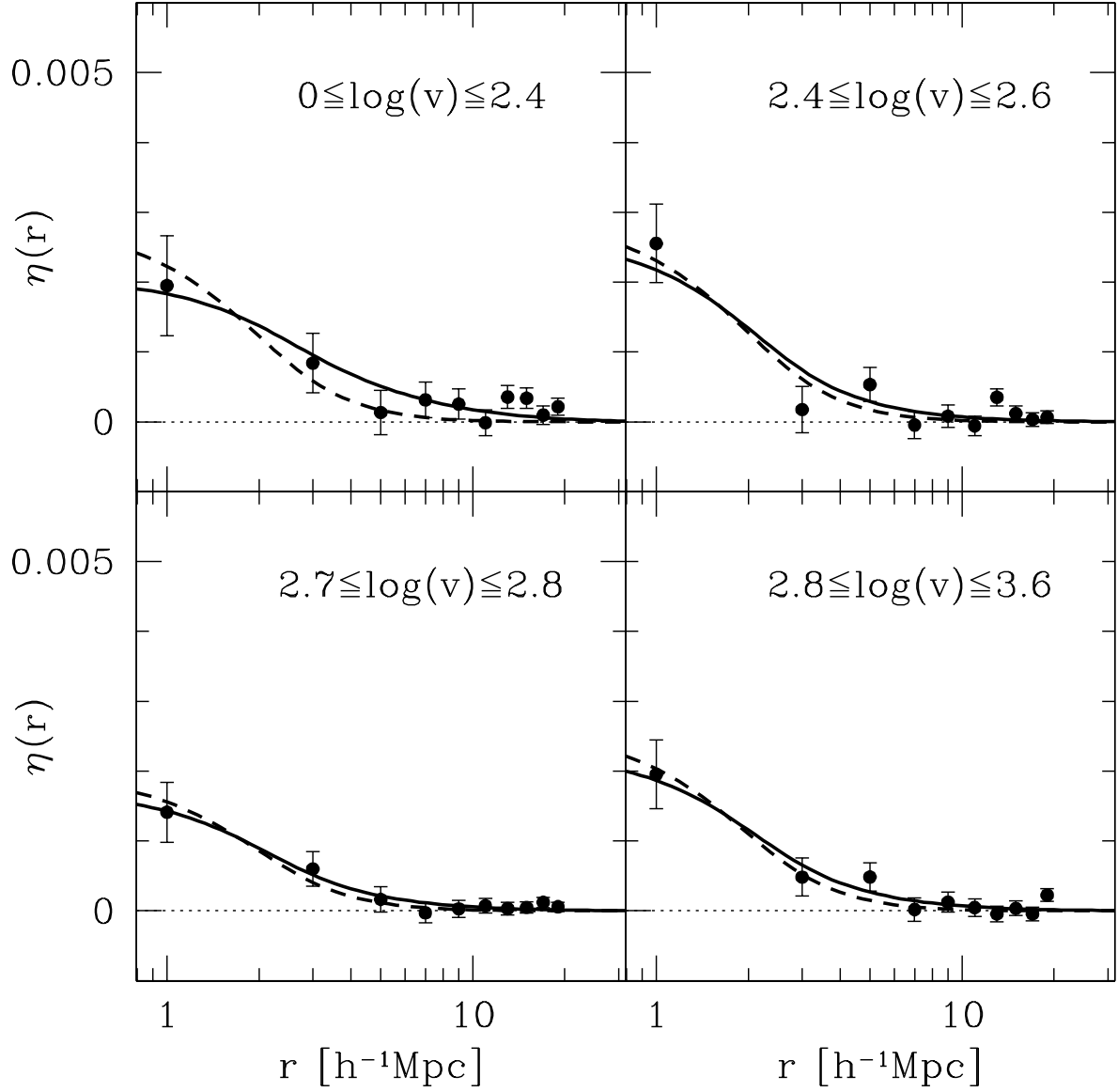}
\caption{Intrinsic halo spin correlations at the four logarithmic bins of 
the velocity magnitude $v$ in unit of km/s.}
\label{fig:v}
 \end{center}
\end{figure}

\begin{figure}
  \begin{center}
   \plotone{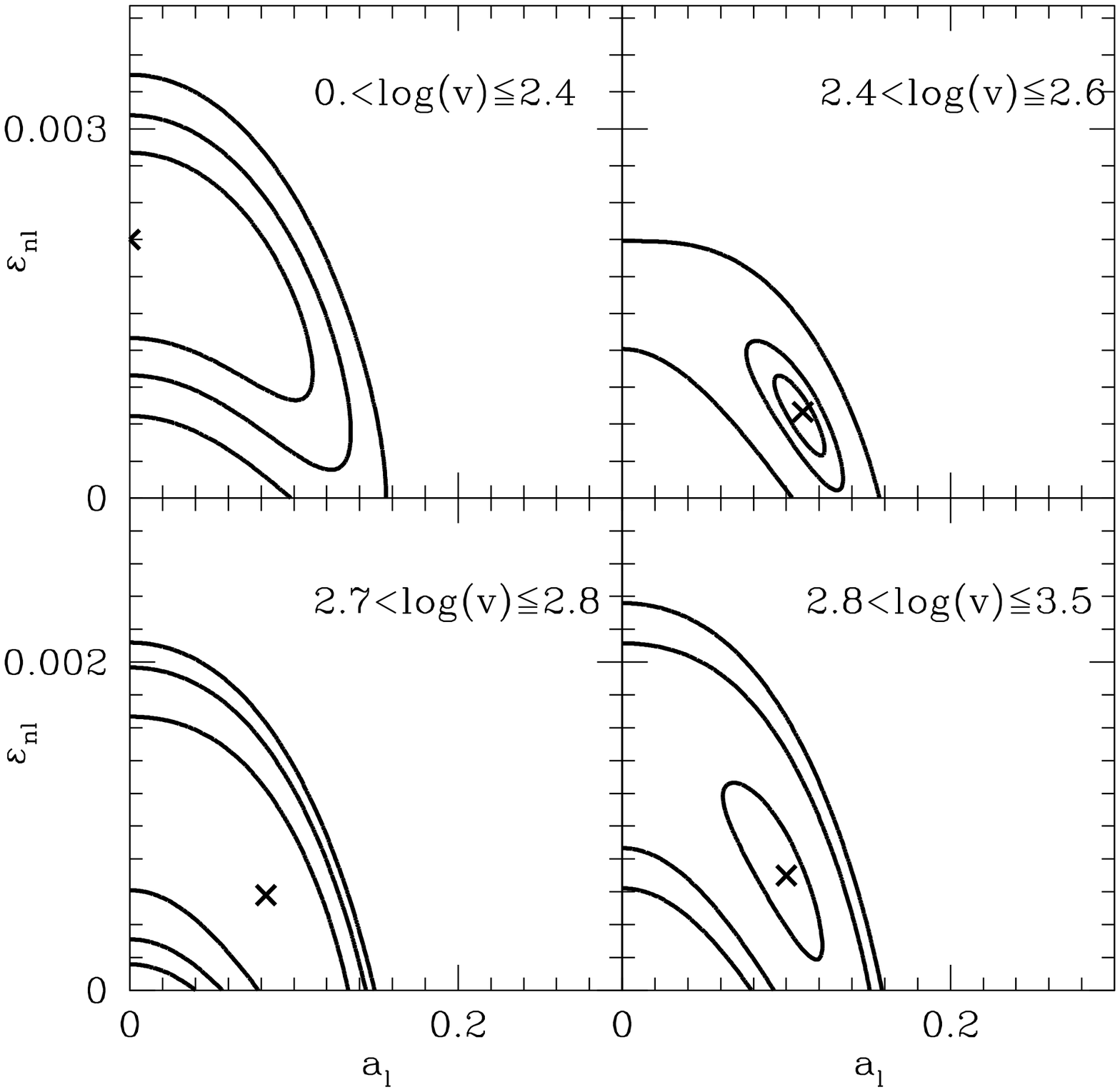}
\caption{$68\%$, $95\%$ and $99\%$ contour plots for $a_{\rm l}$ and 
$\varepsilon_{\rm nl}$ at the four logarithmic bins of $v$.}
\label{fig:cv}
 \end{center}
\end{figure}

We also investigate the dependence of the halo spin-spin correlations 
on the magnitude of halo peculiar velocity $v$ in unit of km/s. The 
correlations $\eta(r)$ are plotted in Fig.\ref{fig:v} at four logarithmic 
bins of $v$, and the $68\%$, $95\%$ and $99\%$ contours for the correlation 
parameters of the nonlinear models are shown in Fig.\ref{fig:cv}.

As shown in Figure \ref{fig:v}, the dependence of the halo spin-spin 
correlations on the velocity magnitude turns out not to be so strong.
The significant nonlinear effect on the halo-halo correlation is found  
only in the first bin of low-$v$ halos (top-left panel), For this bin, 
the linear model fails in fitting the numerical data while the nonlinear 
model gives a better fit, predicting the large-scale correlations. 
As shown in the top-left panel of Fig.\ref{fig:cv}, the value of 
$\varepsilon_{\rm nl}$ for the case of lowest-$v$ halos deviates 
from zero at $95\%$ confidence level.

\section{SUMMARY AND DISCUSSION}

We summarize our results in the following:
\begin{itemize}
\item
To account for the possible nonlinear tidal effect on the intrinsic 
halo alignments, we suggest a new model characterized by two free 
parameters, $a_{\rm l}$ and $\varepsilon_{\rm nl}$. The two parameters
$a_{\rm l}$ and $\varepsilon_{\rm nl}$ represents the strength of the 
linear and the nonlinear tidal effect on the orientations of the halo spins, 
respectively. If $\varepsilon_{\rm nl}=0$, it corresponds to the case 
that the nonlinear tidal effect is negligible and the halo spin-spin 
correlations is a quadratic scaling of the density correlation function 
$\xi(r)$, existent on small scales. While if $\varepsilon_{\rm nl}$ deviates 
from zero, the nonlinear tidal effect is nennegligible and the halo 
spin-spin correlations is a linear scaling of $\xi(r)$, present on large 
distance scales.
\item
Using the halo catalogs from numerical simulations at $z=0,1,0.5$ and $2$, 
we have measured the halo spin-spin correlations and determined the best-fit 
values of $a_{\rm l}$ and $\varepsilon_{\rm nl}$. It is found that the 
values of $a_{\rm l}$ increases with $z$ while $\varepsilon_{\rm nl}$ 
decreases with $z$. 
At $z=1$ and $2$, the values of $\varepsilon_{\rm nl}$ are found to be 
negligibly small, while at $z=0.5$ and $z=0$ $\varepsilon_{\rm nl}$ has 
an order of $10^{-3}$ higher than $5\sigma$. Especially at $z=0$, 
$a_{\rm l}$ is effectively zero and $\varepsilon_{\rm nl}$ deviates from 
zero by a factor more than $10\sigma$, which implies that at present epoch 
the nonlinear tidal effect is dominant inducing halo spin correlations 
on scales as large as $10h^{-1}$Mpc.
\item 
We have investigated the dependence of $a_{\rm l}$ and 
$\varepsilon_{\rm nl}$ on the halo intrinsic properties such as mass ($M$), 
specific angular momentum magnitude ($l$), and velocity magnitude ($v$) at 
$z=0$. It is found that the value of $\varepsilon_{\rm nl}$ depends most 
sensitively on the value of $l$. For those halos with highest $l$, the 
value of $\varepsilon_{\rm nl}$ reaches as high as $5\times 10^{-3}$ 
and the value of $a_{\rm l}$ is effectively zero. 
Regarding the dependence on $M$ and $v$, it is found that 
$\varepsilon_{\rm l}$ is less sensitive. However, for those halos with  
$M \sim (1-2)\times 10^{12}h^{-1}M_{\odot}$ $\log v < 2.5$km/s, the 
value of $\varepsilon_{\rm nl}$ tend to be higher.
\item
We have measured the cross-correlations of the spin axes between halos 
on different mass scales. It is found that on two different galactic mass 
scales ($M_{1} \sim 10^{11}h^{-1}M_{\odot}$ and 
$M_{s} \sim 10^{12}h^{-1}M_{\odot}$, the cross-correlations exist at 
scales as large as $10h^{-1}$Mpc, due to the growth of the non-Gaussianity 
increases cross-correlations of the tidal field on different scales. 
\end{itemize}

Our results have shown that the galaxy intrinsic alignments can exist not only 
for the case that the separation distance between the two galaxies is small 
but also for the case that any galaxy pair separated by distances larger than 
$10h^{-1}$Mpc can still have aligned axes. It also implies preferential E-mode 
contamination of the galaxy intrinsic alignments in weak lensing signals 
\citep{cri-etal02}.

In addition, we have also found quantitatively that the galaxy intrinsic 
alignments could be cross-correlated with the larger-scale tidal field. These 
results imply that the intrinsic galaxy correlations are indeed correlated 
with the surrounding density field on large scales, which will in turn lead 
to correlation between the gravitational lensing shears and the intrinsic 
galaxy alignments, (often called, GI correlations). 

Recently, \citet{hir-etal07} claimed that a signal of the GI correlations has 
been detected from the large low-redshift galaxy surveys 
\citep[see also,][]{man-etal06}. 
According to their claim, the GI correlations exist to a nennegligible level 
for the bright Luminous Red Galaxies (LRGs) and it is zero for the faint blue 
galaxies, mentioning an uncertainty in the estimation of the GI correlation for 
the faint blue galaxies. According to our numerical detection, the nonlinear 
tidal effect which causes the large-scale intrinsic alignments is  strongest 
for the halos with high specific angular momentum magnitude and low peculiar 
velocity, which usually correspond to the galaxies with low surface brightness 
\citep[][and references therein]{jim-etal98} and blue color \citep{dav-etal97}.
Thus, our results indicate that the GI correlation amplitude for the faint blue 
galaxies, may exceed the estimation upper limit, as noted by 
\citet{hir-etal07}. It will be of interest to compare the cross-correlations 
between the spin axes of the blue galaxies and the shapes of the red 
galaxies for the comparison of their correlation scales.

\acknowledgments

The Millennium Run simulation used in this paper was carried out by the Virgo 
Supercomputing Consortium at the Computing Centre of the Max-Planck Society 
in Garching. We thank V. Springel and G. Lemson for useful comments. We also 
thank an anonymous referee for helpful suggestions. J.L. is very grateful to 
the warm hospitality of L. Kofman and the Canadian Institute for Theoretical 
Astrophysics (CITA) where this work was initiated and performed. 
J.L. acknowledges the financial support from the Korea Science and Engineering 
Foundation (KOSEF) grant funded by the Korean Government 
(MOST, NO. R01-2007-000-10246-0).

\clearpage
\begin{deluxetable}{cccccc}
\tablewidth{0pt}
\setlength{\tabcolsep}{5mm}
\tablehead{
$z$ & $N_{h}$ & $\bar{M}_{*}$ & $a\times 10^{2}$ & $a_{\rm l}\times 10^{2}$ & 
$\varepsilon_{\rm nl}\times 10^{3}$ \\ &  &  ($10^{10}h^{-1}M_{\odot}$) & &  & }
\tablecaption{The number of halos, the mean mass, the best-fit values of the 
linear and the nonlinear correlations parameters for the intrinsic halo spin 
correlations at four redshifts from the Millennium Run data.}
\startdata   
$0$ & $1846776$ & $169.7$ & $11.3\pm 0.6$ & $3.9\pm 1.6$  & $1.2\pm 0.1$  \\ 
$0.5$  & $1942263$ & $133.3$ & $9.8\pm 0.7$ & $6.1\pm 1.1$ & $0.7\pm 0.1$  \\ 
$1$ & $1928690$   & $106.7$ & $8.6\pm 0.8$ & $7.9\pm 0.9$& $0.1\pm 0.1$ \\ 
$2$& $1558504$   & $69.6$  &  $10.6\pm 0.9$ & $10.4\pm 0.9$ &$0.0\pm 0.2$  \\ 
\enddata
\label{tab:z}
\end{deluxetable}
\clearpage
\begin{deluxetable}{cccccc}
\tablewidth{0pt}
\setlength{\tabcolsep}{5mm}
\tablehead{$\log (l)$ & $N_{h}$  & $\bar{M}_{*}$ & $a\times 10^{2}$
& $a_{\rm l}\times 10^{2}$ & $\varepsilon_{\rm nl}\times 10^{3}$ \\ &  & 
($10^{10}h^{-1}M_{\odot}$) &  &  &}
\tablecaption{Numerical results for the intrinsic halo spin correlations 
at four bins of the specific angular momentum magnitude $l$.}
\startdata   
$(-4.0, -2.0)$ & $350541$   & $607.1$ & $9.4\pm 2.4$ & $9.4\pm 2.4$  &
$0.0\pm 0.4$  \\ 
$(-2.0, -1.7)$  & $604356$   & $102.7$ & $10.1\pm 2.2$ & $6.6\pm 3.4$ &
$0.7\pm 0.4$ \\ 
$(-1.7,-1.5)$    & $475842$   & $51.0$ & $16.8\pm 2.1$ & $5.4\pm 6.3$   &
$2.7\pm 0.6$ \\ 
$(-1.5, 0.0)$    & $416032$   & $32.5$  &  $20.2\pm 2.2$ & $0.0\pm 9.7$  &
$4.9\pm 0.8$  \\ 
\enddata
\label{tab:a}
\end{deluxetable}
\clearpage
\begin{deluxetable}{cccccc}
\tablewidth{0pt}
\setlength{\tabcolsep}{5mm}
\tablehead{$M_{*}$ & $N_{h}$  & $\bar{M}_{*}$ & $a\times 10^{2}$   
& $a_{\rm l}\times 10^{2}$ & $\varepsilon_{\rm nl}\times 10^{3}$ \\ 
($10^{10}h^{-1}M_{\odot}$) & &  ($10^{10}h^{-1}M_{\odot}$) &  &  &}
\tablecaption{Numerical results for the intrinsic halo spin correlations 
at four different mass ranges.}
\startdata   
$(17.2,38.0)$    & $925011$   & $25.0$ & $14.1\pm 1.6$ & $8.5\pm 2.6$  &
$1.3\pm 0.4$  \\ 
$(38.0, 85.0)$  & $469647$   & $55.7$ & $14.1\pm 2.3$ & $10.6\pm 3.2$ &
$1.0\pm 0.6$   \\ 
$(85.0,190.0)$    & $233724$   & $124.6$ & $20.8\pm 2.5$ & $15.0\pm 3.5$   &
$2.4\pm 0.9$ \\ 
$(190.0,10^{5})$    & $218248$   & $1022.5$  &  $15.5\pm 1.9$ & 
$13.0\pm 2.2$  & $0.8\pm 0.6$  \\ 
\enddata
\label{tab:m}
\end{deluxetable}

\clearpage
\clearpage
\clearpage
\clearpage
\clearpage
\clearpage
 \begin{figure}
  \begin{center}
   \plotone{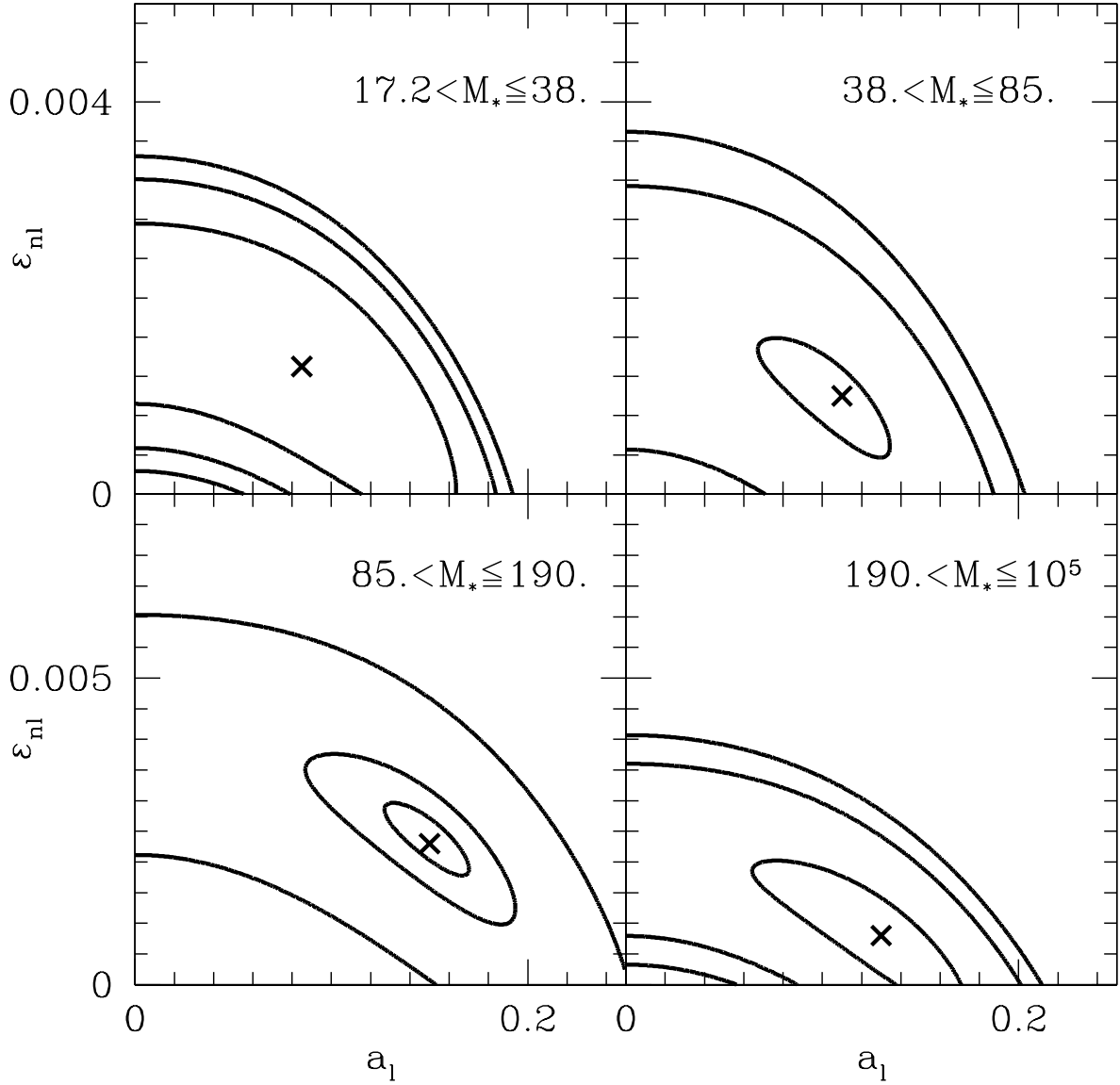}
\caption{$68\%$, $95\%$ and $99\%$ contour plots for  $a_{\rm l}$ and 
$\varepsilon_{\rm nl}$ at the four bins of halo mass $M_{*}$.}
\label{fig:cm}
 \end{center}
\end{figure}
\clearpage
\clearpage
 \begin{figure}
  \begin{center}
   \plotone{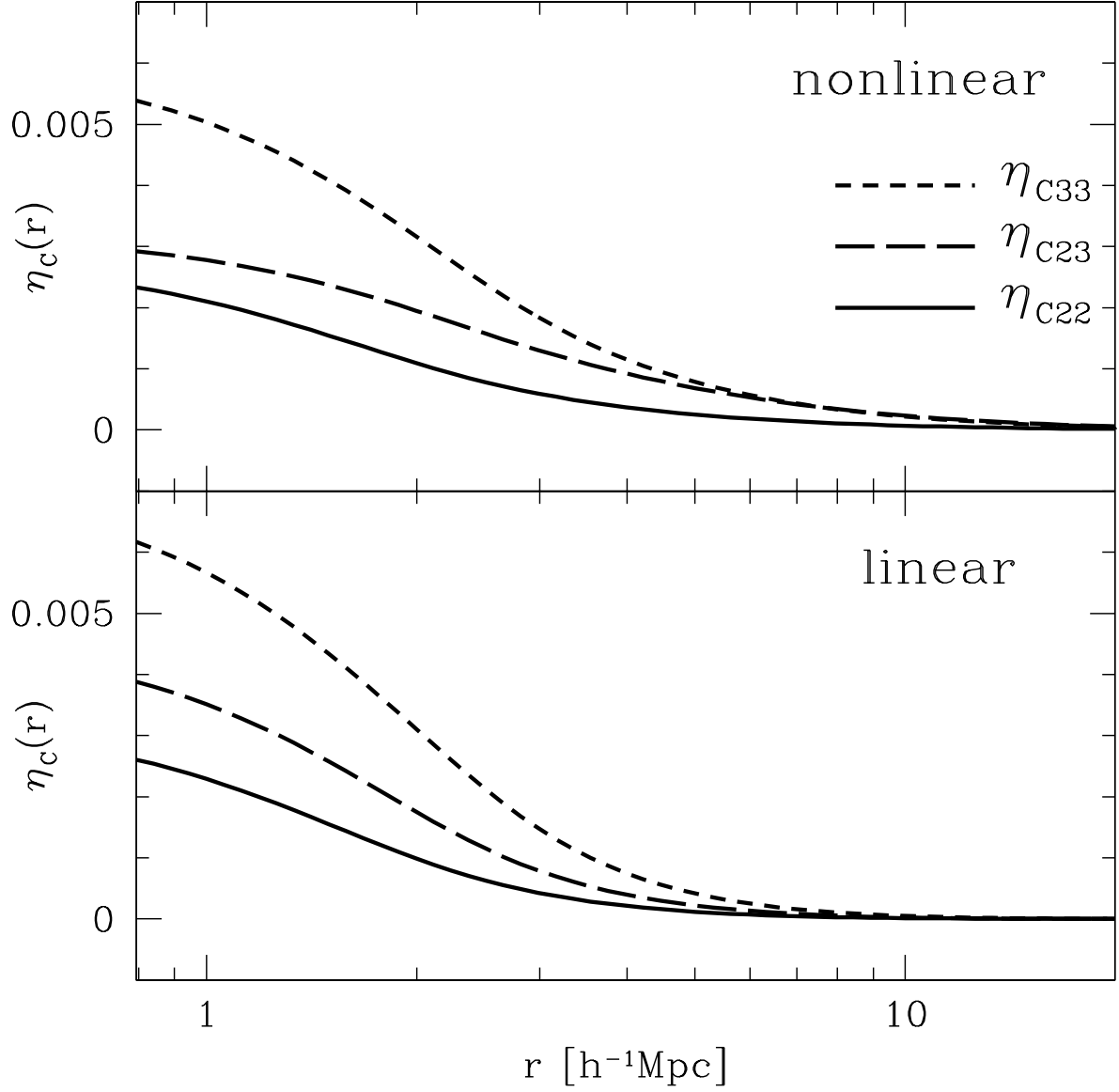}
\caption{Comparison of the spin correlation $\eta_{ii}(r)$ between halos 
in the same mass bins with the spin cross-correlation $\eta_{Cij}(r)$ 
between halos in different mass bins. }
\label{fig:tb}
 \end{center}
\end{figure}
\clearpage
\clearpage

\end{document}